\documentclass[a4paper,12pt]{article}
\usepackage{amssymb}
\usepackage{amsmath}
\usepackage{epsfig}
\usepackage{stmaryrd}
\usepackage{graphics}

\usepackage{latexsym}
\usepackage{rotating}

\title{\bf Exact one-periodic and two-periodic wave
 solutions to Hirota bilinear equations in $2+1$ dimensions}
\date{}

\author{
Wen-Xiu Ma\thanks{Email: {\tt mawx@cas.usf.edu}}
\\
{\small Department of Mathematics and Statistics, University of South Florida, Tampa, FL 33620-5700, USA}
\\
{\small Department of Mathematics, Zhejiang Normal University, Jinhua 321004, P. R. China}
\vspace {2mm}
\\
Ruguang Zhou\thanks{Email: {\tt rgzhou@public.xz.js.cn}}
\\
{\small School of Mathematical Science,
  Xuzhou Normal
University} \\ {\small Xuzhou 221116, P. R. China} \vspace{2mm}
\\
Liang Gao\thanks{Email: {\tt gaoliang@mail.nwpu.edu.cn}} \\ {\small Department of Applied
Mathematics, Northwestern Polytechnical University}\\ {\small Xi'an 710072, P. R. China}
}

\begin{document}

\maketitle

\numberwithin{equation}{section}

\oddsidemargin=0mm \evensidemargin=0mm \baselineskip=18.2pt
\parindent 20pt

\newcommand \Z {\mathbb{Z}}
\newcommand \R {\mathbb{R}}
\newcommand \D {\displaystyle}

\begin{abstract}
Riemann theta functions are used to construct one-periodic and two-periodic wave solutions to a class of
$(2+1)$-dimensional Hirota bilinear equations. The basis for the involved
solution analysis is the Hirota bilinear formulation, and the particular dependence of the equations on independent variables
guarantees the existence of
one-periodic and two-periodic
wave solutions involving an arbitrary purely imaginary Riemann matrix. The resulting theory is applied to two nonlinear equations possessing Hirota
bilinear forms: $u_t+u_{xxy}-3uu_y-3u_xv=0$ and $
u_t+u_{xxxxy}-(5u_{xx}v+10u_{xy}u-15u^2v)_x=0$ where $v_x=u_y$, thereby yielding their
one-periodic and two-periodic wave solutions describing one dimensional propagation of waves.

\vskip 2mm



\end{abstract}

\section{Introduction}

It is always important to search for exact solutions to nonlinear differential equations.
Different approaches, particularly in soliton theory, provide many tools for constructing
explicit and exact solutions. Various kinds of exact solutions such as solitons,
positons, complexitons, solitonoffs and dromions have been presented for nonlinear
integrable equations \cite{AblowitzC-book1991}-\cite{Gao}. Successful methods include the
inverse scattering transform \cite{AblowitzC-book1991}, the Darboux transformation
\cite{MatveevS-book1991}, Hirota direct method \cite{Hirota-book2004}, and
algebro-geometrical approach \cite{BelokolosBEIM-book1994}.

The algebro-geometrical approach presents quasi-periodic or algebro-geometric solutions
to many soliton equations, which contain the KdV equation, the sine-Gordon equation and
the nonlinear Schr\"odinger equation. In recent years, such an approach have been applied
to many (2+1)-dimensional nonlinear integrable equations
\cite{CaoWG-JMP1999}-\cite{GengD-PA2003}. Nonlinearization of Lax pairs
\cite{Cao-SCA1990}-\cite{MaZ-ANZIAMJ2002} plays a crucial role in connecting the resulting algebro-geometric solutions with Liouville integrable Hamiltonian systems.
The approach, however, needs Lax pair representations and involves complicated calculus
on Riemann surfaces.

On the other hand, the Hirota direct method provides a powerful way to derive soliton
solutions to nonlinear integrable equations and its basis is the Hirota bilinear formulation
\cite{Hirota-book2004}. Once the corresponding bilinear forms are obtained, multi-soliton
solutions and rational solutions to nonlinear differential equations can be computed in a
quite systematic way \cite{Hirota-book2004}, even through Wronskian, Casoratian or
Pfaffian determinants \cite{HirotaO-JPSJ1991}-\cite{MaHL-NA2008}. It is based on Hirota
bilinear forms that Nakamura presented an approach to multi-periodic wave
solutions of nonlinear integrable equations \cite{Nakamura-JPSC197980}, together with applications to various typical examples of integrable equations. Such a method of solution does not need any Lax pairs and their
induced Riemann surfaces for the considered equations. The presented multi-periodic
solutions can be reduced to soliton solutions under asymptotic limits
\cite{Matruno-book,Zhang-JPA2007}. The advantage of the method is that it only relies on
the existence of Hirota bilinear forms. Moreover, all parameters appearing in Riemann
matrices are completely arbitrary, whereas algebro-geometric solutions involve specific
Riemann constants, which are usually difficult to compute.

In this paper, motivated by Nakamura's idea \cite{Nakamura-JPSC197980}, we would like to
use Riemann theta functions to generate one-periodic and two-periodic wave solutions to a particular
class of (2+1)-dimensional Hirota bilinear equations, and the corresponding
solution analysis will be made to guarantee the existence of one-periodic and two-periodic wave
solutions to the selected class of (2+1)-dimensional nonlinear equations. As illustrative
examples of the resulting theory, we will discuss two nonlinear equations possessing Hirota
bilinear forms:
 \[u_t+u_{xxy}-3uu_y-3u_xv=0\ \ \textrm{and}\ \
u_t+u_{xxxxy}-(5u_{xx}v+10u_{xy}u-15u^2v)_x=0,\] where $v_x=u_y$, and their one-periodic
and two-periodic wave solutions involving an arbitrary
purely imaginary Riemann matrix
will be explicitly presented.


\section{Existence of one-periodic and two-periodic wave solutions}

Let us consider an evolution equation in $2+1$ dimensions:
\begin{equation}
u_t=K(u,u_x,u_y,\cdots),\label{eq:2+1dsolitonquation:ma174}\end{equation}
where $t\in \R $ is the time variable and $x,y\in \R $ are the space variables. We assume that under a transformation
\begin{equation} u=u_0-2(\ln f(x,y,t))_{xx},\end{equation}
where $u_0$ is a special solution to (\ref{eq:2+1dsolitonquation:ma174}), the evolution equation
\eqref{eq:2+1dsolitonquation:ma174} can be transformed into a Hirota bilinear equation
\begin{equation}
F(D_x,D_y,D_t)f\cdot f=0, \label{eq:2+1dgeneralbilinearequation:ma174}
\end{equation}
where $F$ is a polynomial in the three variables. Here and below, the Hirota bilinear
differential operators \cite{Hirota-book2004} are defined by
\begin{equation}
\begin{array} {l}
D_x^pD_y^qD_t^r f(x,y,t)\cdot g(x,y,t) \vspace{2mm}\\ =(\partial_x-\partial_{x'})^p
(\partial_y-\partial_{y'})^q(\partial_t-\partial_{t'})^r f(x,y,t)
g(x',y',t')|_{x'=x,y'=y,t'=t},\end{array}
\end{equation}
 where $p,q,r$ are
non-negative integers. We will focus on a particular class of Hirota bilinear equations
in $2+1$ dimensions:
\begin{equation}
F(D_x,D_y,D_t)f\cdot f=(D_tP(D_x)+D_yQ(D_x)+R(D_x))f\cdot f=0,
\label{eq:2+1dbilinearequation:ma174}
\end{equation}
where $P$ and $Q$ are nonzero odd polynomials and $R$ is a nonzero even polynomial,
namely, $P,\, Q$ and $R$ are nonzero polynomials and satisfy
\begin{equation}
P(-z)=-P(z),\ Q(-z)=-Q(z), \ R(-z)=R(z).\label{eq:propertyofPQR:ma174}\end{equation}

When the Hirota operators act on exponential
  functions, the following derivative formula holds:
  \begin{equation}
  D_x^pD_y^qD_t^r e^{\eta_1}\cdot
  e^{\eta_2}=(k_1-k_2)^p(l_1-l_2)^q(\omega_1-\omega_2)^r
  e^{\eta _1+\eta_2},\end{equation}
where $\eta_j=k_jx+l_jy+\omega_jt+\eta _{j0},\ j=1,2$, with $k_j,l_j,\omega_j,\eta_{j0}$
being constants. More generally, we have
\begin{equation} G(D_x,D_y,D_t)e^{\eta_1}\cdot
  e^{\eta_2}=G(k_1-k_2,l_1-l_2,\omega_1-\omega_2)
  e^{\eta_1+\eta_2},\label{eq:propertyforDoperator:ma174} \end{equation}
where $G$ is a polynomial in the three variables. This derivative formula will be a
crucial key to our success in generating one-periodic and two-periodic wave solutions.

We would like to consider the multi-dimensional special Riemann theta function solution
\cite{RauchF-book1974}:
\begin{equation}
f=f(x,y,t)=\sum_{n\in \Z^N}e^{ 2\pi i\langle \eta ,n\rangle+\pi i\langle \tau n,n\rangle
},\label{eq:defoff:ma174}
\end{equation} where
 $\langle \cdot,\cdot \rangle$ is the standard
inner product of $\R ^N$, $n=(n_1,\cdots,n_N)^T,$ $ \eta =(\eta _1, \cdots, \eta _N)^T$
with $\eta_j=k_jx+l_jy+\omega_j t+\eta_{j0}$, and $\tau=(\tau_{pq})_{N\times N}$ is a
symmetric matrix whose imaginary part is positive definite (i.e., ${\rm Im}\,\tau>0$).
Base on \eqref{eq:propertyforDoperator:ma174}, we can compute $G(D_x,D_t,\cdots)f\cdot f$
generally for such a Riemann theta function $f$ \cite{HirotaI-JPSJ1981}, but we will make
direct computations to provide a complete solution process and capture more of special
solution structures.

\subsection{One-periodic wave solutions}

Let us first consider the case of $N=1$. Then the Riemann theta function in
\eqref{eq:defoff:ma174} becomes
\begin{equation}
f=f(x,y,t)=\sum_{n=-\infty}^{\infty} e^{2\pi i n\eta +\pi i
n^2\tau},\label{eq:defof1df:ma174}
\end{equation}
where $\rm{Im}\,\tau >0$ and $ \eta =kx+ly+\omega t +\eta _0$ with $k,l,\omega,\eta_0$
being real constants.

Based on the derivative formula (\ref{eq:propertyforDoperator:ma174}),
 we can compute that
\begin{eqnarray*}
&&F(D_x,D_y,D_t) f\cdot f=F(D_x,D_y,D_t)\sum_{n=-\infty}^{\infty}e^{2\pi i n\eta +\pi i
n^2\tau}\cdot\sum_{m=-\infty}^{\infty} e^{2\pi i m\eta +\pi i m^2\tau}\\
&&=\sum_{n=-\infty}^{\infty}\sum_{m=-\infty}^{\infty}F(D_x,D_y,D_t)e^{2\pi i n\eta +\pi i
n^2\tau}\cdot e^{2\pi i m\eta +\pi i m^2\tau}\\
&&=\sum_{n=-\infty}^{\infty}\sum_{m=-\infty}^{\infty}F(2\pi i(n-m)k,2\pi i(n-m)l,2\pi
i(n-m)\omega)e^{2\pi i(n+m)\eta +\pi i ( n^2+m^2)\tau}\\
&&=\sum_{m'=-\infty}^{\infty}\bigl\{ \sum_{n=-\infty}^{\infty}F(2\pi i(2n-m')k,2\pi
i(2n-m')l,2\pi i(2n-m')\omega)e^{\pi i[(n^2+(n-m')^2]\tau}\bigr\}e^{2\pi i m'\eta }\\
&&=\sum_{m'=-\infty}^{\infty}{\tilde F}(m')e^{2\pi i m'\eta },
\end{eqnarray*}
where the new summation $m'=m+n$ has been introduced and $\tilde F(m')$ is defined by
\begin{equation}
{\tilde F}(m')=\sum_{n=-\infty}^{\infty}F(2\pi i(2n-m')k,2\pi i(2n-m')l,2\pi
i(2n-m')\omega)e^{\pi i[n^2+(n-m')^2]\tau}.
\end{equation}
Shifting index $n $ by introducing $n'=n-1$, we have
\begin{eqnarray*}
{\tilde F}(m')&=&\sum_{n=-\infty}^{\infty}F(2\pi i(2n-m')k,2\pi i(2n-m')l,2\pi
i(2n-m')\omega)e^{\pi i[n^2+(n-m')^2]\tau}\\ &=&\sum_{n'=-\infty}^{\infty}F(2\pi
i[2n'-(m'-2)]k,2\pi i[2n'-(m'-2)]l,2\pi i[2n'-(m'-2)]\omega)\\ &&\ \ \ \ \  \  \ \times
e^{\pi i\{n'^2+[n'-(m'-2)]^2\}\tau} e^{2\pi i(m'-1)\tau} \\ &=&{\tilde F}(m'-2)e^{2\pi i
(m'-1)\tau},\ m'\in \Z. \end{eqnarray*} It then follows that if ${\tilde F}(0)={\tilde
F}(1)=0$, then ${\tilde F}(m')=0$ for all $ m'\in \Z$.

Noticing the specific form of the equation (\ref{eq:2+1dbilinearequation:ma174}),
one-periodic wave solutions can be obtained, if we require
\begin{equation}\left\{\begin{array}{l}
\D \tilde F(0)=\sum_{n=-\infty}^\infty [4n\pi i \omega P(4n \pi i k) +4n\pi i l Q(4n \pi
i k)  +R (4n \pi i k)  ]e^{2n^2\pi i \tau }=0,\vspace{2mm}\\ \D \tilde
F(1)=\sum_{n=-\infty}^\infty [2(2n-1)\pi i \omega P(2(2n-1) \pi i k) +2(2n-1)\pi i l
Q(2(2n-1) \pi i k) \vspace{2mm}\\ \qquad \quad \qquad \  +R (2(2n-1) \pi i k)
]e^{(2n^2-2n+1)\pi i \tau }=0.\end{array}\right.
\label{eq:resultinglinearsystemOf1periodicwave:ma174}
\end{equation}
Upon introducing
\begin{equation}\left\{ \begin{array}{l}
\D a_{11}(k)=\sum_{n=-\infty}^\infty  4n \pi i P(4n \pi i k) e^{2n^2\pi i \tau
},\vspace{2mm}\\
  \D a_{12}(k)=\sum_{n=-\infty}^\infty 4n \pi i
Q(4n \pi i k) e^{2n^2\pi i \tau },\vspace{2mm}\\ \D a_{21}(k)=\sum_{n=-\infty}^\infty
2(2n-1) \pi i P(2(2n-1) \pi i k) e^{(2n^2-2n+1)\pi i \tau },\vspace{2mm}\\ \D
a_{22}(k)=\sum_{n=-\infty}^\infty  2(2n-1) \pi i Q(2(2n-1) \pi i k) e^{(2n^2-2n+1)\pi i
\tau },\vspace{2mm}\\
\end{array}\right.
\end{equation}
and
\begin{equation}\left\{ \begin{array}{l}
\D b_{1}(k)=-\sum_{n=-\infty}^\infty  R(4n \pi i k) e^{2n^2\pi i \tau },\vspace{2mm}\\
  \D b_{2}(k)=-\sum_{n=-\infty}^\infty  R(2(2n-1) \pi i k)
e^{(2n^2-2n+1)\pi i \tau },
\end{array}\right.\label{eq:defofb[1-2](k):ma174}
\end{equation}
the linear system \eqref{eq:resultinglinearsystemOf1periodicwave:ma174}  of $\omega$ and
$l$ can be compactly written as
\begin{equation}\begin{array}{l}
a_{11}(k)\omega +a_{12}(k)l =b_1(k), \ a_{21}(k)\omega +a_{22}(k)l =b_2(k).
\end{array}
\label{eq:simpleformOfresultinglinearsystemOf1periodicwave:ma174}
\end{equation}

We will see that there are a lot of choices for the angular wave number $k$. In order to
generate real solutions $(\omega,l)$ to the system
\eqref{eq:simpleformOfresultinglinearsystemOf1periodicwave:ma174}, we assume that
\begin{equation} {\rm Re}\,\tau =0. \label{eq:RealPartoftauIsZero:ma174} \end{equation}
 The determinant of the
coefficient matrix $A(k)=(a_{rs}(k))_{2\times 2}$ is a polynomial in $k$, and so, if
$\det (A(k))\not \equiv 0$ (this condition will be satisfied in our concrete examples),
then
 \begin{equation} A_0:=\{k\in \R \, | \det (A(k))= 0\}\end{equation} is either
an empty set or a finite set. This guarantees the existence of real solutions
$(\omega,l)$ to the system
\eqref{eq:simpleformOfresultinglinearsystemOf1periodicwave:ma174} at least for $k\not\in
A_0$. About nonzero solutions, we can have the following analysis.

 If deg$(R)=0$, i.e., $R=c$, where $c$ is a
nonzero real constant, then it follows
from \eqref{eq:defofb[1-2](k):ma174} that $b(k)$ does not depend on $k$ and
\[ b(k)=(b_1(k),b_2(k))^T \ne 0, \] and so, there
is the unique nonzero solution of $(\omega, l)$ to the system
\eqref{eq:simpleformOfresultinglinearsystemOf1periodicwave:ma174} for $k\not \in A_0$.

If deg$(R)\ge 2$, then \begin{equation}
 B_0:=\{k\in \mathbb{R}\, | (b_1(k))^2+b_2(k))^2= 0\} \end{equation}
 is either an empty set or a finite set, since each of $b_1(k)$ and
 $b_2(k)$ is a polynomial in $k$ of degree $\deg (R)$.
Therefore, there is the unique nonzero solution of $(\omega, l)$ to the system
\eqref{eq:simpleformOfresultinglinearsystemOf1periodicwave:ma174} for $k\not \in A_0\cup
B_0$.

\subsection{Two-periodic wave solutions}

Let us second consider the case of $N=2$ and the corresponding two-periodic wave solutions.
Similarly, based on the
derivative formula (\ref{eq:propertyforDoperator:ma174}) and introducing $m'=n+m$,
 we can have
\begin{eqnarray*}
&& F(D_x,D_y,D_t)f\cdot f=\sum_{m,n\in \Z^2} F(D_x,D_y,D_t)e^{2\pi i\langle \eta
,n\rangle +\pi i\langle \tau n,n\rangle }\cdot e^{2\pi i\langle \eta ,m \rangle +\pi
i\langle \tau m,m\rangle }\\ &&=\sum_{m,n\in \Z^2}F(2\pi i\langle n-m,k\rangle , 2\pi
i\langle n-m,l\rangle , 2\pi i\langle n-m,\omega\rangle )e^{2\pi i\langle \eta,n+m\rangle
+\pi i (\langle \tau m,m\rangle +\langle \tau n, n\rangle )}\\ &&{=}\sum_{m'\in
\Z^2}\sum_{n\in \Z ^2} F(2\pi i\langle 2n-m',k\rangle , 2\pi i\langle 2n-m',l\rangle ,
2\pi i\langle 2n-m',\omega\rangle )\\ &&\ \ \ \ \ \ \ \ \ \ \ \ \ \quad \times e^{\pi
i(\langle \tau (n-m'),n-m'\rangle +\langle \tau n, n\rangle )} e^{2\pi i\langle
\eta,m'\rangle }\\ &&= \sum_{m'\in \Z^2}{\tilde F}(m_1', m_2') e^{2\pi i \langle \eta
,m'\rangle },
\end{eqnarray*}
where
 $\tilde F(m_1',m_2')={\tilde F}(m')$ is defined by
\begin{equation}
{\tilde F}(m_1',m_2')= \sum_{n\in \Z ^2} F(2\pi i\langle 2n-m',k\rangle , 2\pi i\langle
2n-m',l\rangle , 2\pi i\langle 2n-m',\omega\rangle ) e^{\pi i(\langle \tau
(n-m'),n-m'\rangle +\langle \tau n, n\rangle )} .
\end{equation}
Shifting index $n $ as $n'=n-e_{r}$ with $r=1$ or $r=2$, where $e_1=(1,0)^T$ and
$e_2=(0,1)^T$, we can compute that
\begin{eqnarray*}
&&{\tilde F}(m_1', m_2')={\tilde F}(m')={\tilde F}(m'-2e_r)e^{2\pi i (\langle \tau
(m'-2e_r),e_r \rangle + \langle \tau e_r,e_r \rangle )} \vspace{2mm} \\
&&=\left\{\begin{matrix}{\tilde F}(m_1'-2,m_2')e^{2\pi i(m_1'-1)\tau_{11}+2\pi
im_2'\tau_{12}},\ \ r=1,\vspace{2mm} \cr {\tilde F}(m_1',m_2'-2) e^{2\pi
i(m_2'-1)\tau_{22}+2\pi im_1'\tau_{12}},\ \ r=2,
\end{matrix}\right.
\end{eqnarray*}
where $\tau =(\tau_{pq})_{2\times 2}$. It now follows that if
\begin{equation}
{\tilde F}(0,0)={\tilde F}(0,1)={\tilde F}(1,0)={\tilde
F}(1,1)=0,\label{eq:resultinglinearsystemOf2periodicwave:ma174}
\end{equation}
 then ${\tilde F}(m_1',m_2')=0$ for all $ m_1',m_2'\in \Z$.

For our selected equation (\ref{eq:2+1dbilinearequation:ma174}), we have
\begin{equation}\begin{array}{l}
\D \tilde F(m_1,m_2)=\sum_{n\in \Z ^2} [2\pi i \langle 2n-m,\omega \rangle P(2\pi i
\langle 2n-m,k\rangle ) \vspace{2mm}\\ \qquad \qquad \qquad  \qquad    \D
 +2\pi i \langle
2n-m,l\rangle  Q(2\pi i \langle 2n-m,k\rangle ) \vspace{2mm}\\ \qquad \qquad \qquad
\qquad     \D  +R (2\pi i \langle 2n-m,k\rangle ) ]e^{\pi i ( \langle \tau (n-m),n-m
\rangle +\langle \tau n,n\rangle )},\end{array} \qquad
\label{eq:resultingTildeFOf2periodicwave:ma174}
\end{equation}
where we set \begin{equation} m=(m_1,m_2)^T,\ n=(n_1,n_2)^T, \ k=(k_1,k_2)^T,\
l=(l_1,l_2)^T,\ \omega =(\omega _1,\omega _2)^T .\end{equation} For simplicity, define
\begin{equation} \theta _r(n)= e^{ \pi i( \langle \tau
(n-m^{(r)}),n-m^{(r)}\rangle + \langle \tau n,n\rangle ) }, \ 1\le r\le 4,\end{equation}
 where $m^{(r)}=(m_1^{(r)},m_2^{(r)})^T$, $1\le
r\le 4$, are given by
 \[  m^{(1)}=(0,0)^T,\ m^{(2)}=(0,1)^T,\
m^{(3)}=(1,0)^T,\ m^{(4)}=(1,1)^T .\] Then, upon introducing
\begin{equation}\left\{\begin{array}{l}
\D a_{rs}(k)=\sum_{n_1,n_2=-\infty}^\infty  2 \pi i (2n_s-m_s^{(r)})P(2 \pi i  \langle
2n-m^{(r)},k\rangle   ) \theta _{r}(n) ,\vspace{2mm}\\ \D
a_{r,s+2}(k)=\sum_{n_1,n_2=-\infty}^\infty  2 \pi i (2n_s-m_s^{(r)})Q(2 \pi i
 \langle 2n-m^{(r)},k\rangle
) \theta _{r}(n) ,
\end{array}\right.
\end{equation}
where $1\le r\le 4 $ and $1\le s\le 2$, and
\begin{equation}
\D b_r(k)=-\sum_{n_1,n_2=-\infty}^\infty  R(2 \pi i  \langle 2n-m^{(r)},k\rangle ) \theta
_r(n),\ 1\le r\le 4, \label{eq:defofb[1-4](k):ma174}
\end{equation}
 the linear system
\eqref{eq:resultinglinearsystemOf2periodicwave:ma174} of $(\omega,l)$ can be compactly
written as
\begin{equation}A(k) \left [\begin{array}{c}
\omega _{1} \vspace{2mm}\\ \omega_2 \vspace{2mm}\\ l_1\vspace{2mm}\\ l_2
\end{array}\right] =b(k)=
\left [\begin{array}{c} b_{1}(k) \vspace{2mm}\\ b_2(k) \vspace{2mm}\\
b_3(k)\vspace{2mm}\\ b_4(k)
\end{array}\right],
\label{eq:simpleformOfresultinglinearsystemOf2periodicwave:ma174}
\end{equation}
where $A(k)=(a_{rs}(k))_{4\times 4}$. If $\tau $ is purely imaginary, i.e., it satisfies
\eqref{eq:RealPartoftauIsZero:ma174}, then $A(k)$ and $b(k)$ are real, due to our
assumption on the polynomials $P,Q,R$. Note that if $\det (A(k))\not \equiv 0$ (this
condition will be satisfied in our concrete examples), then
\begin{equation} A_0:=\{k\in \R ^2\, | \det (A(k))=0\}\end{equation} is either an empty
set or
a finite set.

 Now if deg$(R)=0$, i.e.,
$R=c$, where $c$ is a nonzero real constant, then it follows from \eqref{eq:defofb[1-4](k):ma174} that $b(k)$ does not depend on
$k$ and $ b(k)\ne  0,$ and so, there is the unique nonzero solution of $(\omega_1,
\omega_2,l_1,l_2)$ to the system
\eqref{eq:simpleformOfresultinglinearsystemOf2periodicwave:ma174} for $k\not \in A_0$.

If deg$(R)\ge 2$, then \begin{equation}
 B_0:=\{k\in \mathbb{R}^2\, | \sum_{r=0}^4 (b_r(k))^2= 0\} \end{equation}
 is either an empty set or a finite set, since each of $b_r(k),\ 1\le r\le
 4$,
 is a polynomial in $k_1$ and $k_2$ of degree $\deg R$.
Therefore, there is the unique nonzero solution of $(\omega_1, \omega_2,l_1,l_2)$ to the
system \eqref{eq:simpleformOfresultinglinearsystemOf2periodicwave:ma174} for $k\not \in
A_0\cup B_0$.

\section{Two illustrative examples}

Let us illustrate our idea of generating one-periodic and two-periodic wave solutions through two
particular Hirota bilinear equations. The first example is
\begin{equation}
u_t+u_{xxy}-3uu_y-3u_xv=0,\ v_x=u_y,\label{eq:1stexample:ma174}
\end{equation}
in the physical field. This nonlinear equation is related to the breaking soliton
equation \cite{Carogelo-LNC1978}:
\[
u_t+u_{xxy}-4uu_y-2u_x \partial_x ^{-1} u_y=0,
\] and it can be
transformed into
\begin{equation}
(D_tD_x+D_yD_x^3+c)f\cdot f=0,\label{eq:bleof1stexample:ma174}
\end{equation}
where $c$ can be an arbitrary function of $y$ and $t$, under the transformation
\begin{equation}u=-2(\ln f)_{xx},\ v=-2(\ln f)_{xy}.
\label{eq:variabletransformationof1stexample:ma174}\end{equation} Actually, we have
\[ u_t+u_{xxy}-3uu_y-3u_xv = -\bigl( \frac {(D_tD_x+D_yD_x^3)f\cdot f}{f^2} \bigr)_x. \]
The second example is
\begin{equation}
u_t+u_{xxxxy}-(5u_{xx}v+10u_{xy}u-15u^2v)_x=0,\ v_x=u_y,\label{eq:2ndexample:ma174}
\end{equation} in the physical field.
This nonlinear equation can be transformed into
\begin{equation}
(D_tD_x+D_yD_x^5+c)f\cdot f=0, \label{eq:bleof2ndexample:ma174}
\end{equation}
where $c$ can be an arbitrary function of $y$ and $t$, under the same transformation
\eqref{eq:variabletransformationof1stexample:ma174}. Similarly, we have
\[ u_t+u_{xxxxy}-(5u_{xx}v+10u_{xy}u-15u^2v)_x= -\bigl( \frac {(D_tD_x+D_yD_x^5)f\cdot
f}{f^2} \bigr)_x. \]
The involved arbitrary function $c$ of $y$ and $t$ shows the diversity of solutions to (2+1)-dimensional
differential equations.

To generate one-periodic and two-periodic wave solutions by the solution method in the last section, we need to assume that the above function $c$ is constant, based on which
the angular wave number $l$
(or the pair of angular wave numbers $l_1$ and $l_2$) and the frequency
$\omega $ (or the pair of frequencies $\omega_1$ and $\omega _2$) are constant and thus the derivative formula \eqref{eq:propertyforDoperator:ma174} will hold.
Obviously, we have
\begin{equation} P(z)=z,\ Q(z)=z^3, \ R(z)=c,\end{equation}
for the equation \eqref{eq:1stexample:ma174} and
\begin{equation} P(z)=z,\ Q(z)=z^5, \ R(z)=c,\end{equation}
for the equation \eqref{eq:2ndexample:ma174}. The polynomials $P$ and $ Q$ defined above
are odd and the polynomials $R$ defined above are even, and so, the property
\eqref{eq:propertyofPQR:ma174} is satisfied. The determinants of the corresponding coefficient matrices of the
linear systems \eqref{eq:simpleformOfresultinglinearsystemOf1periodicwave:ma174} and
\eqref{eq:simpleformOfresultinglinearsystemOf2periodicwave:ma174}
 are not
identically equal to zero, namely,
\[\det (A(k))\not \equiv 0\  \ \textrm{and}\  \ \det (A(k_1,k_2))\not \equiv 0\]
in the two examples. For instance, in the case of one-periodic wave solutions, we have
 \begin{equation} \det (A(k))=a k^4  \ \ \textrm{or}\ \  \det (A(k))=bk^6,\end{equation}
  where \[
 \begin{array}{l}
\D a= -256\pi ^6\sum_{n=-\infty}^\infty n^2 e^{2n^2\pi i\tau } \sum_{n=-\infty}^\infty
(2n-1)^4 e^{(2n^2-2n+1)\pi i\tau } \vspace{2mm}\\ \D \qquad +1024 \pi ^6
\sum_{n=-\infty}^\infty n^4 e^{2n^2\pi i\tau }  \sum_{n=-\infty}^\infty (2n-1)^2
e^{(2n^2-2n+1)\pi i\tau }
 ,\vspace{2mm}\\
\D
 b=1024\pi ^8\sum_{n=-\infty}^\infty n^2 e^{2n^2\pi i\tau }
 \sum_{n=-\infty}^\infty (2n-1)^6 e^{(2n^2-2n+1)\pi i\tau } \vspace{2mm}\\
\D \qquad -16384 \pi ^8 \sum_{n=-\infty}^\infty n^6 e^{2n^2\pi i\tau }
\sum_{n=-\infty}^\infty (2n-1)^2 e^{(2n^2-2n+1)\pi i\tau } .
 \end{array}
 \]
A direct computation by Maple 11 with Digits = 30 shows that
\[ \begin{array}{l}
a|_{\tau=0.1i}\approx 4563.212514,\  a|_{\tau=0.2i}\approx 140396.7042,\
a|_{\tau=0.5i}\approx 25831.08621, \vspace{2mm}\\
 b|_{\tau=0.1i}\approx 11012599.24,\  b|_{\tau=0.2i}\approx 28544399.95,\
 b|_{\tau=0.5i}\approx
 -4884657.870,
 \end{array}
\]
which are all nonzero. Generally, our general analysis made before is valid for the two
equations \eqref{eq:bleof1stexample:ma174} and \eqref{eq:bleof2ndexample:ma174}, and so,
one-periodic and two-periodic wave solutions to
the two $(2+1)$-dimensional
nonlinear equations possessing Hirota
bilinear forms, \eqref{eq:1stexample:ma174}
and \eqref{eq:2ndexample:ma174},
can be
computed explicitly.

\section{Conclusion and remarks}

The Riemann theta functions have been used to generate one-periodic and two-periodic wave solutions of a
particular class of (2+1)-dimensional Hirota bilinear equations, and the
corresponding solution analysis has been made to guarantee the existence of
such multi-periodic wave solutions. Two illustrative examples:
 \[u_t+u_{xxy}-3uu_y-3u_xv=0\ \ \textrm{and}\ \
u_t+u_{xxxxy}-(5u_{xx}v+10u_{xy}u-15u^2v)_x=0,\]
 where $v_x=u_y$,
have been discussed in details, along with their one-periodic and two-periodic wave
solutions involving an arbitrary purely imaginary Riemann matrix.

Our solution analysis
provides a way to construct one-periodic and two-periodic wave solutions to (2+1)-dimensional nonlinear
differential equations. It allows different angular wave numbers $k$ (or
different pairs of angular wave numbers
$k_1$ and $k_2$), but the angular wave number $l$ (or the pair of angular wave numbers $l_1$ and $l_2$) and the frequency $\omega $ (or the pair of frequencies $\omega_1$ and $\omega_2$) are determined in terms of each angular wave number $k$ (or each pair of angular wave numbers $k_1$ and $k_2$) and hence the obtained solutions describe one-dimensional propagation of waves.

We also remark that the proposed approach can be applied to other nonlinear differential
equations.
For example, the following combined equation with the Sawada-Kotera vector field:
\[
u_t+u_{xxy}-3uu_y-3u_xv+u_{xxxxx}-15(uu_{xx}-u^3)_x=0,\ v_x=u_y,
\]
can be analyzed similarly. Under the transformation
\eqref{eq:variabletransformationof1stexample:ma174}, this equation can be put into the
following bilinear equation:
 \[
(D_tD_x+D_yD_x^3+D_x^6+c)f\cdot f=0,\] where $c$ can be an arbitrary function of $y$ and $t$.
The corresponding polynomials $P,Q,R$ read
\begin{equation} P(z)=z,\ Q(z)=z^3, \ R(z)=z^6+c,\end{equation}
where $c$ is assumed to be constant.
Therefore, the same analysis on one-periodic and two-periodic wave solutions will work for this equation
as well. On the other hand, soliton solutions to the equations
\eqref{eq:1stexample:ma174} and \eqref{eq:2ndexample:ma174} can be computed by using
Hirota's direct method. For example, one soliton solutions to the equations
\eqref{eq:1stexample:ma174} and \eqref{eq:2ndexample:ma174} are determined by
\[f=1+e^{\pm k^3t+kx\mp ky}\ \ \textrm{and}\ \ f=1+e^{\pm k^5t+kx\mp ky},\ \ k
\textrm{ - arbitrary const.}, \]
respectively. This can also be verified by using \eqref{eq:propertyforDoperator:ma174}.
It should be, however, interesting to establish any relations between soliton solutions
and multi-periodic wave solutions.

It is our hope that our analysis on one-periodic and two-periodic wave solutions made for the particularly selected class of Hirota bilinear equations could help to better
understand the diversity and integrability of nonlinear differential equations.


\vskip 4mm

{\leftline{\normalsize {\bf Acknowledgments}}}


{\small The work was supported in part by the Established Researcher Grant of the
University of South Florida, the CAS faculty development grant of the University of South
Florida, Chunhui Plan of the Ministry of Education of China, Wang Kuancheng foundation,
the National Natural Science Foundation of China (Grant Nos. 10871165, 10332030, 10472091,
10502042), and the Doctorate Foundation of Northwestern Polytechnical University (Grant
No. CX200616).}


\begin{thebibliography}{99}
\footnotesize
 \bibitem{AblowitzC-book1991} M. J. Ablowitz and  P. A. Clarkson, {\it Solitons,
     Nonlinear Evolution
           Equations and Inverse Scattering} (Cambridge University Press, Cambridge,
           1991).

\bibitem{MatveevS-book1991}  V. B. Matveev and M. A. Salle, {\it Darboux Transformation
    and Solitons}
            (Springer,  Berlin, 1991).

\bibitem{Hirota-book2004} R. Hirota, {\it Direct Methods in Soliton Theory} (Springer,
    Berlin, 2004).

\bibitem{BelokolosBEIM-book1994}  E. Belokolos, A. Bobenko, V. Enol'skij, A, Its and V.
    Matveev,
            {\it Algebro-Geometrical Approach to Nonlinear Integrable
              Equations} (Springer, Berlin, 1994).

\bibitem{Boiti} M. Boiti, J. Jp. Leon, L. Martina and F. Pempinelli,
    {\it Phys. Lett. A} {\bf 132}, 432 (1988).

\bibitem{Hietarinta-PLA1990}  J. Hietarinta,
    {\it Phys. Lett. A} {\bf 149}, 113 (1990).

\bibitem{Lou}
 S. Y. Lou, X. Y. Tang, X. M. Qian, C. L. Chen, J. Lin and S. L. Zhang,
{\it Modern Phys.
Lett. B} {\bf 16}, 1075 (2002).

\bibitem{MaY-TAMS2005}W. X. Ma and Y. You,
    {\it Trans. Amer. Math. Soc.} {\bf 357}, 1753
    (2005).

\bibitem{Gao} L. Gao, W. Xu, Y. N. Tang and G. F. Meng,
    {\it Phys. Lett. A} {\bf 366}, 411 (2007).

\bibitem{CaoWG-JMP1999}  C. W. Cao, Y. T. Wu and X. G. Geng, {\it J. Math. Phys.} {\bf
    40}, 3948 (1999).

\bibitem{CaoGW-JMP2002}  C. W. Cao,  X. G. Geng and H. Y. Wang, {\it J. Math. Phys.} {\bf
    43}, 621 (2002).

\bibitem{Zhou} R. G. Zhou,
    {\it Nuovo Cimento B} {\bf 117}, 925 (2002).

 \bibitem{GengD-PA2003}
  X. G. Geng and H. H. Dai,
{\it Phys. A} {\bf
319}, 270 (2003).

\bibitem{Cao-SCA1990}C. W. Cao, {\it Sci. China Ser. A} {\bf 33}, 528 (1990).

\bibitem{MaS-PLA1994}W. X. Ma and W. Strampp,
  {\it Phys. Lett. A} {\bf 185},
    277 (1994).

\bibitem{MaG-book2001}W. X. Ma and X. G. Geng,
in: {\it B\"acklund and Darboux transformations -
    The Geometry of Solitons} (Halifax, NS, 1999), 313-323, CRM Proc. Lecture Notes {\bf
    29} (Amer. Math. Soc., Providence, RI, 2001).

\bibitem{MaZ-ANZIAMJ2002} W. X. Ma and Y. B. Zeng,
{\it ANZIAM J.} {\bf 44}, 129 (2002).

\bibitem{HirotaO-JPSJ1991} R. Hirota and Y. Ohta,  {\it J. Phys. Soc. Jpn.} {\bf 60}, 798
    (1991).

\bibitem{Ma-PLA2002}W. X. Ma,
    {\it Phys. Lett. A} {\bf 301}, 35 (2002).

\bibitem{MaM-PA2004} W. X. Ma and K. Maruno,
    {\it Phys. A} {\bf 343}, 219 (2004).

\bibitem{ZhaoLH-JPSJ2004} J. X. Zhao, C. X. Li and X. B. Hu,  {\it J. Phys. Soc. Jpn.}
    {\bf 73}, 1159 (2004).

\bibitem{HuLNY-JPA2005} X. B. Hu, C. X. Li, J. J. C. Nimmo and G. F. Yu, {\it J. Phys A:
    Math. Gen.} {\bf 38}, 195 (2005).

\bibitem{LiMLZ-IP2007}
C. X. Li, W. X. Ma, X. J. Liu and Y. B. Zeng,
{\it Inverse Problems} {\bf 23}, 279 (2007).

\bibitem{MaHL-NA2008}W. X. Ma, J. S. He and C. X. Li,
A second Wronskian formulation of the Boussinesq equation,
{\it Nonlinear Anal. Theor. Meth. Appl.},
Available online 5 October 2008.

\bibitem{Nakamura-JPSC197980}A. Nakamura, {\it J. Phys. Soc. Jpn.} {\bf 47}, 1701 (1979);
    {\bf 48}, 1365 (1980).

\bibitem{Matruno-book}Y. Matsuno, {\it Bilinear Transformation Method} (Academic Press,
    Orlando, 1984).

\bibitem{Zhang-JPA2007} Y. Zhang, L. Y. Ye, Y. N. Lv and H. Q. Zhao,
{\it J. Phys. A: Math. Theor.} {\bf 40}, 5539
    (2007). 

\bibitem{RauchF-book1974}
 H. E. Rauch and H. M. Farkas,  {\it Theta Functions with Applications to Riemann
 Surfaces}
 (The Williams $\&$ Wilkins Co., Baltimore, Md., 1974).

\bibitem{HirotaI-JPSJ1981}R. Hirota and M. Ito, {\it J. Phys. Soc. Jpn.} {\bf 50}, 338
    (1981).

\bibitem{Carogelo-LNC1978} F. Calogero and A. Degasperis, {\it Nuovo Cimento B} {\bf 31},
    201 (1977).

\end{thebibliography}
\end{document}